\documentclass[12pt]{article}
\usepackage{graphics}
\usepackage{bm}
\usepackage{graphicx}
\usepackage{amsmath}
\usepackage{amssymb}
\usepackage{amstext}
\usepackage{epstopdf}
\usepackage{amscd}
\usepackage{amsfonts}
\usepackage{appendix}
\usepackage{color}
\usepackage{wasysym}
\usepackage{textcomp}
\DeclareMathAlphabet{\EuFrak}{U}{euf}{m}{n}
\DeclareMathAlphabet{\EuScript}{U}{eus}{m}{n}

\newcommand{\nd}{\noindent}

\newcommand{\be}{\begin{equation}}
\newcommand{\ee}{\end{equation}}
\newcommand{\ben}{\begin{eqnarray}}
\newcommand{\een}{\end{eqnarray}}

\title{{\bf Dimensionally regularized Tsallis' Statistical
Mechanics and two-body Newton's gravitation}}
\author{{J. D. Zamora$^{1,4}$, M. C. Rocca$^{1,2,4}$,A. Plastino$^{1,4,5}$,
G. L. Ferri $^3$}, \\
\small{$^1$ Departamento de F\'{\i}sica,
Universidad Nacional de La Plata,}\\
\small{$^2$ Departamento de Matem\'{a}tica,
Universidad Nacional de La Plata,}\\
\small{$^3$Fac. de C. Exactas-National University La Pampa,} \\
\small{Peru y Uruguay, Santa Rosa, La Pampa, Argentina}\\
\small{$^4$ Consejo Nacional de Investigaciones Cient\'{\i}ficas
y Tecnol\'{o}gicas}\\
\small{(IFLP-CCT-CONICET)-C. C. 727, 1900 La Plata -
Argentina}\\\small{$^5$  SThAR - EPFL, Lausanne, Switzerland}}

\date{\today}

\begin{document}

\maketitle

\begin{abstract}

\nd Typical Tsallis' statistical mechanics' quantifiers like the partition 
function and the mean energy exhibit
poles.  We are speaking of the partition function 
${\cal Z}$ and the mean
energy  $<{\cal U}>$.  The poles appear for distinctive values of 
Tsallis' characteristic real 
parameter $q$, at a numerable set of
rational numbers of the $q-$line. These poles are dealt with dimensional regularization resources.
 The physical effects
of these poles on the specific heats
are studied here for the two-body classical gravitation potential. 


\end{abstract}

\newpage

\renewcommand{\theequation}{\arabic{section}.\arabic{equation}}

\section{Introduction}

Tsallis'  information measure $S_T$ generalizes Shannon's  one and is considered 
a very important statistical quantifier \cite{tsallis88}, \cite{tsallisbook}, \cite{[44]},   
 \cite{[5]},   \cite{[6]},  \cite{[7]},  \cite{[8]}, \cite{[9]},  \cite{[10]},
 \cite{[11]}, \cite{[12]}, \cite{[13]}, \cite{[14]},  etc.
 $S_T$ reads 
\cite{tsallisbook}  
\begin{equation}
\label{eq2.1} S_T=\frac {1-\int\limits_{{\cal M}}
P^qd\mu} {q-1}
\end{equation}
The dimensional regularization of $S_T$ was discussed in \cite{12} 
by  investigating the 
poles that emerge in computing the partition function and the mean energy (${\cal Z}$, 
and  $<{\cal U}>$, respectively), ascertaining  the physical significance of these poles for the harmonic oscillator (HO). In this paper we do the same for the classical gravitation potential.  To do so we appeal to the dimensional regularization approach of Bollini and  Giambiagi
\cite{tq1}, \cite{tq2,tq4,tq5}, generalized as explained in \cite{tp1}.   
Dimensional regularization constitutes one of the most important theoretical physics' advances of the 20-th century's second half. It is used in several  of its disciplines  \cite{dr1}-\cite{dr54}. In particular, 
we heavily rely on the article \cite{epl},   a quite useful   prerequisite.
More details about this topic are given in Appendix A of this paper.
\vskip 3mm 
\nd The statistical mechanics of systems ruled by gravity is connected to aspects of  condensed matter physics, fluid mechanics, re-normalization
group, etc. It constitutes a challenge with  regards  to basic foundations.
 Associated notions may encounter application in variegated  areas of astrophysics and cosmology.
 Among many statistical  works  of this kind one could recommend, for instance,  
  \cite{[1], [2], [3],[4]}. Here we address our subject via dimensional regularization.
\setcounter{equation}{0}

\section{The theory for $q>1$}

Consider the two-body Newton's gravity and its Tsallis' statistical mechanics. For the partition function one 
deals with 
\begin{equation}
\label{ep2.1}
{\cal Z}_{\nu}=\int\limits_{{\cal M}}\left[1+(1-q)\beta\left(\frac {p^2} {2m}-
\frac {GmM} {r}\right)\right]_+^{\frac {1} {q-1}}d^\nu xd^\nu p
\end{equation}
For effecting the integration process one uses hyper-spherical coordinates and two integrals, each in  $\nu$ dimensions. 
Ones is left with just two radial coordinates (one in $r-$ space and the other in $p-$ space) and 
$2(\nu -1)$ angles. Since the argument of the brackets must be positive, one has 
\[{\cal Z}\nu=\left[\frac {2\pi^{\frac {\nu} {2}}} {\Gamma\left(
\frac {\nu} {2}\right)}\right]^2
[\beta(q-1)]^{\frac {1} {q-1}}
\int\limits_0^\infty r^{\nu-1}dr\otimes\]
\begin{equation}
\label{ep2.2}
\int\limits_0^{{\sqrt{2m\left(\frac {1} {\beta(q-1)}+
\frac {GmM} {r}\right)}}}
p^{\nu-1}
\left[\frac {1} {\beta(q-1)}+
\frac {GmM} {r}-\frac {p^2} {2m}\right]^{\frac {1} {q-1}}dp,
\end{equation}
where by $[]_+$ we mean that one considers only that $p$ region where the bracket is positive, which 
 entails that the  $p$ integration runs from $0$ till ${\sqrt{2m\left(\frac {1} {\beta(q-1)}+
\frac {GmM} {r}\right)}}$. This is called the Tsallis' cut-off.
The two integrals above can be evaluated by recourse to 
Euler's Beta function ${\cal B}$ \cite{gr}).
We give the result of the first one, let us call it $I_1$.
 \[I_1= \int\limits_0^{{\sqrt{2m\left(\frac {1} {\beta(q-1)}+
\frac {GmM} {r}\right)}}}
p^{\nu-1}
\left[\frac {1} {\beta(q-1)}+
\frac {GmM} {r}-\frac {p^2} {2m}\right]^{\frac {1} {q-1}}dp=\]
\begin{equation}
\label{er1}
\left[\frac {1} {\beta(q-1)}+
\frac {GmM} {r}\right]^{\frac {\nu} {2}+\frac {1} {q-1}}
{\cal B}\left(\frac {\nu} {2}, \frac {1} {q-1}+1\right).
\end{equation}
Accordingly,
\begin{equation}
\label{ep2.3}
{\cal Z}_\nu=\frac {2\pi^\nu(2m)^{\frac {\nu} {2}}} {\left[\Gamma\left(
\frac {\nu} {2}\right)\right]^2}
[\beta(q-1)]^{\frac {\nu} {2}}(GmM)^\nu
{\cal B}\left(\frac {\nu} {2},\frac {1} {q-1}+1\right)
{\cal B}\left(\frac {\nu} {2}+\frac {1} {1-q},-\nu\right)
\end{equation}
From (\ref{ep2.3}) one gathers that poles appear for any dimension $\nu$, 
 $\nu=3$ included. Thus, appeal to dimensional regularization (DR) is mandatory. To this effect we will use the   DR-generalization given in \cite{tp1} of 
Bollini - Giambiagi's originl DR-technique. \vskip 3mm 

\nd To proceed further we face now 
\begin{equation}
\label{ep2.4}
{\cal Z}_\nu<{\cal U}>_\nu=\int\limits_{{\cal M}}\left[1+(1-q)\beta\left(\frac {p^2} {2m}-
\frac {GmM} {r}\right)\right]_+^{\frac {1} {q-1}}
\left(\frac {p^2} {2m}-\frac {GmM} {r}\right)d^\nu xd^\nu p,
\end{equation}
and 
\[{\cal Z}_\nu<{\cal U}>_\nu=\left[\frac {2\pi^{\frac {\nu} {2}}} {\Gamma\left(
\frac {\nu} {2}\right)}\right]^2
[\beta(q-1)]^{\frac {1} {q-1}}\left\{
\int\limits_0^\infty r^{\nu-1}\right.dr\otimes\]
\[\int\limits_0^{\sqrt{2m\left(\frac {1} {\beta(q-1)}+
\frac {GmM} {r}\right)}}
\frac {p^{\nu+1}} {2m}
\left[\frac {1} {\beta(q-1)}+
\frac {GmM} {r}-\frac {p^2} {2m}\right]^{\frac {1} {q-1}}dp-\]
\begin{equation}
\label{ep2.5}
\left.GmM\int\limits_0^\infty r^{\nu-2}dr
\int\limits_0^{\sqrt{2m\left(\frac {1} {\beta(q-1)}+
\frac {GmM} {r}\right)}}
p^{\nu-1}
\left[\frac {1} {\beta(q-1)}+
\frac {GmM} {r}-\frac {p^2} {2m}\right]^{\frac {1} {q-1}}dp\right\}.
\end{equation}
Beta functions were invented by Euler and we not give an explicit form of them here because
they appear in almost all fields of physics. For more details
see the Appendix. Via the Beta function one finds 
\[<{\cal U}>_\nu=\frac {2\pi^\nu(2m)^{\frac {\nu} {2}}} {{\cal Z}_\nu\left[\Gamma\left(
\frac {\nu} {2}\right)\right]^2}
[\beta(q-1)]^{\frac {\nu} {2}-1}(GmM)^\nu\left[
{\cal B}\left(\frac {\nu} {2}+1,\frac {1} {q-1}+1\right)\otimes\right.\]
\begin{equation}
\label{ep2.6}
\left.{\cal B}\left(\frac {\nu} {2}+\frac {1} {1-q},1-\nu\right)-
{\cal B}\left(\frac {\nu} {2},\frac {1} {q-1}+1\right)
{\cal B}\left(\frac {\nu} {2}+\frac {1} {1-q}-1,1-\nu\right)\right.
\end{equation}

\setcounter{equation}{0}

\section{The theory for $q<1$}   

The treatment becomes more complicated in this instance. 
From (\ref{ep2.1}) we find
\[{\cal Z}_\nu=\left[\frac {2\pi^{\frac {\nu} {2}}} {\Gamma\left(
\frac {\nu} {2}\right)}\right]^2
[\beta(1-q)]^{\frac {1} {q-1}}\left\{
\int\limits_0^{GmM\beta(1-q)} r^{\nu-1}dr\right.\otimes\]
\[\int\limits_{\sqrt{2m\left(\frac {GmM} {r}-
\frac {1} {\beta(q-1)}\right)}}^\infty
p^{\nu-1}
\left[\frac {p^2} {2m}-
\frac {GmM} {r}+\frac {1} {\beta(1-q)}\right]_+^{\frac {1} {q-1}}dp+\]
\begin{equation}
\label{ep3.1}
\left.\int\limits_{GmM\beta(1-q)}^\infty r^{\nu-1}dr
\int\limits_0^\infty
p^{\nu-1}
\left[\frac {p^2} {2m}-
\frac {GmM} {r}+\frac {1} {\beta(1-q)}\right]_+^{\frac {1} {q-1}}dp\right\}.
\end{equation}
We deal with four integrals that are evaluated by appeal to Beta funnctions. 
\[{\cal Z}_\nu=\frac {2\pi^\nu(2m)^{\frac {\nu} {2}}} {\left[\Gamma\left(
\frac {\nu} {2}\right)\right]^2}
[\beta(q-1)]^{\frac {\nu} {2}}(GmM)^\nu\left[
{\cal B}\left(\frac {\nu} {2},\frac {1} {1-q}-\frac {\nu} {2}\right)\otimes\right.\]
\[{\cal B}\left(-\nu,\frac {\nu} {2}-\frac {1} {1-q}+1\right)
+{\cal B}\left(\frac {1} {1-q}-\frac {\nu} {2},\frac {1} {q-1}+1\right)\otimes\]
\begin{equation}
\label{ep3.2}
\left.{\cal B}\left(\frac {\nu} {2}+\frac {1} {q-1}+1,\frac {\nu} {2}+\frac {1} {1-q}\right).
\right]
\end{equation}
Looking for the mean energy we deal, from (\ref{ep2.4}), with 
\[{\cal Z}_\nu<{\cal U}>_\nu=\left[\frac {2\pi^{\frac {\nu} {2}}} {\Gamma\left(
\frac {\nu} {2}\right)}\right]^2
[\beta(1-q)]^{\frac {1} {q-1}}\left\{
\int\limits_0^{GmM\beta(1-q)} r^{\nu-1}dr\right.\otimes\]
\[\int\limits_{\sqrt{2m\left(\frac {GmM} {r}-
\frac {1} {\beta(q-1)}\right)}}^\infty
\frac {p^{\nu+1}} {2m}
\left[\frac {p^2} {2m}-
\frac {GmM} {r}+\frac {1} {\beta(1-q)}\right]_+^{\frac {1} {q-1}}dp+\]
\[\int\limits_{GmM\beta(1-q)}^\infty r^{\nu-1}dr
\int\limits_0^\infty
\frac {p^{\nu+1}} {2m}
\left[\frac {p^2} {2m}-
\frac {GmM} {r}+\frac {1} {\beta(1-q)}\right]_+^{\frac {1} {q-1}}dp-\]
\[GmM
\int\limits_0^{GmM\beta(1-q)} r^{\nu-2}dr
\int\limits_{\sqrt{2m\left(\frac {GmM} {r}-
\frac {1} {\beta(q-1)}\right)}}^\infty
p^{\nu-1}
\left[\frac {p^2} {2m}-
\frac {GmM} {r}+\frac {1} {\beta(1-q)}\right]_+^{\frac {1} {q-1}}dp-\]
\begin{equation}
\label{ep3.3}
GmM\left.\int\limits_{GmM\beta(1-q)}^\infty r^{\nu-1}dr
\int\limits_0^\infty
p^{\nu-1}
\left[\frac {p^2} {2m}-
\frac {GmM} {r}+\frac {1} {\beta(1-q)}\right]_+^{\frac {1} {q-1}}dp\right\},
\end{equation}
involving eight integrals. Beta functions are again needed. We have

\[<{\cal U}>_\nu=\frac {2\pi^\nu(2m)^{\frac {\nu} {2}}} {{\cal Z}_\nu\left[\Gamma\left(
\frac {\nu} {2}\right)\right]^2}
[\beta(q-1)]^{\frac {\nu} {2}-1}(GmM)^\nu\left[
{\cal B}\left(\frac {1} {1-q}-\frac {\nu} {2}-1,\frac {1} {q-1}+1\right)\otimes\right.\]
\[{\cal B}\left(\frac {\nu} {2}+\frac {1} {q-1}+2,\frac {\nu} {2}+\frac {1} {1-q}-1\right)+
{\cal B}\left(\frac {\nu} {2}+1,\frac {1} {1-q}-\frac {\nu} {2}-1\right)\otimes\]
\[{\cal B}\left(-\nu,\frac {\nu} {2}-\frac {1} {1-q}+2\right)-
{\cal B}\left(\frac {1} {1-q}-\frac {\nu} {2},\frac {1} {q-1}+1\right)\otimes\]
\[{\cal B}\left(\frac {\nu} {2}+\frac {1} {q-1}+1,\frac {\nu} {2}+\frac {1} {1-q}-1\right)-
{\cal B}\left(\frac {\nu} {2},\frac {1} {1-q}-\frac {\nu} {2}\right)\otimes\]
\begin{equation}
\label{ep3.4}
\left.{\cal B}\left(1-\nu,\frac {\nu} {2}-\frac {1} {1-q}+1\right)\right].
\end{equation}
Dimensional regularization is needed.

\section{The divergences of the theory}

\setcounter{equation}{0}

From  (\ref{ep3.4}) we gather that the mean energy can not be regularized for some  $q$ values, 
those such that 
\begin{equation}
\label{ep4.1}
1+\frac {1} {q-1}=-n\;\;{\rm for} \;\;n=0,1,2,3,......,
\end{equation}
DR can be attempted whenever
\begin{equation}
\label{ep4.2}
1+\frac {1} {q-1}\neq-n\;\;{\rm for} \;\;n=0,1,2,3,......,
\end{equation}
or, equivalently, for
\begin{equation}
\label{ep4.3}
q\neq\frac {1} {2},\frac {2} {3},\frac {3}
{4},\frac {4} {5},......, \frac {\nu-2} {\nu-1},\frac {\nu-1}
{\nu}, \frac {\nu} {\nu+1},.......
\end{equation}
We emphasize that here we have  $q<1$.

\section{The three-dimensional scenario for $q>1$}

\setcounter{equation}{0}

Let  us deal with the  
$q=\frac {3} {2}$ instance. 
 We go back to (\ref{ep2.3}). The idea it to work out the dimensional regularization task and its  corresponding Laurent expansion. We have
\[{\cal Z}_\nu=-\frac {8\pi^2(\beta G^2m^3M^2)^{\frac {3} {2}}} {3(\nu-3)}+
\frac {4\pi^2} {3}(\beta G^2m^3M^2)^{\frac {3} {2}}\left[
\frac {23} {3}-
\ln\left(16\pi^2\beta G^2m^3M^2\right)\right]+\]
\begin{equation}
\label{ep5.1}
\sum\limits_{s=1}^{\infty}a_s(\nu-3)^s.
\end{equation} 
 In the present case $\nu-3$, the independent term
in the ${\cal Z}$-Laurent expansion yields the physical value of the series.
Thus, 
\begin{equation}
\label{ep5.2}
{\cal Z}=\frac {4\pi^2} {3}(\beta G^2m^3M^2)^{\frac {3} {2}}\left[
\frac {23} {3}-
\ln\left(16\pi^2\beta G^2m^3M^2\right)\right].
\end{equation}
Since ${\cal Z}$ must be positive, one faces a temperature-lower bound
\begin{equation}
\label{eq5.3}
T>\frac {e^{-\frac {23} {3}}} {k_B}16\pi^2G^2m^3M^2.
\end{equation}
\vskip 3mm 
\nd 
Similarly,  from (\ref{ep2.6}), we  have for $<{\cal U}>$
\[{\cal Z}<{\cal U}>_\nu=\frac {32\pi^2(\beta G^2m^3M^2)^{\frac {3} {2}}} {3(\nu-3)}-
\frac {32\pi^2} {3}(\beta G^2m^3M^2)^{\frac {3} {2}}\left[
3-\ln\left(\pi^2\beta G^2m^3M^2\right)\right]+\]
\begin{equation}
\label{ep5.4}
\sum\limits_{s=1}^{\infty}a_s(\nu-3)^s.
\end{equation}
Accordingly, 
\begin{equation}
\label{ep5.5}
{\cal Z}<{\cal U}>=
\frac {32\pi^2} {\beta}(\beta G^2m^3M^2)^{\frac {3} {2}}\left[
\ln\left(\pi^2\beta G^2m^3M^2\right)-3-2\boldsymbol{C}\right],
\end{equation}
and
\begin{equation}
\label{ep5.6}
<{\cal U}>=
\frac {8[\ln\left(\pi^2\beta G^2m^3M^2\right)-3-2\boldsymbol{C}]}
{\beta[\frac {23} {3}-\ln\left(16\pi^2\beta G^2m^3M^2\right)}].
\end{equation}
where $\boldsymbol{C}$ is the Euler's constant \cite{grad}.

\section{The three-dimensional scenario for $q<1$}

\setcounter{equation}{0}

Consider $q=\frac {1} {3}$. The concomitant  Laurent expansion derived from 
(\ref{ep3.2}) is 
\begin{equation}
\label{ep6.1}
{\cal Z}=\frac {4} {9\pi}(4\pi^2\beta G^2m^3M^2)^{\frac {3} {2}}\left[
\boldsymbol{C}+8\ln 2+\frac {10} {3}-
\ln\left(\frac {4\pi^2\beta G^2m^3M^2} {3}\right)\right].
\end{equation}
Positivity of  ${\cal Z}$ leads us again to a $T$-lower bound: 
\begin{equation}
\label{eq6.2}
T>\frac {e^{-(\boldsymbol{C}+8\ln 2+\frac {10} {3})}} {3k_B}
4\pi^2G^2m^3M^2.
\end{equation}
For $<{\cal U}>$ we deduce, from  (\ref{ep3.4}),
\begin{equation}
\label{ep6.3}
{\cal Z}<{\cal U}>=
\frac {1} {2\pi}(4\pi^2\beta G^2m^3M^2)^{\frac {3} {2}}\left[8+3\ln 3-
3\ln\left(\pi^2\beta G^2m^3M^2\right)-\ln 16-5\boldsymbol{C}\right],
\end{equation}
and
\begin{equation}
\label{ep6.4}
<{\cal U}>=\frac {9} {8\beta}\;
\frac {8+3\ln 3-3\ln\left(4\pi^2\beta G^2m^3M^2\right)
-\ln 16-5\boldsymbol{C}]}
{\boldsymbol{C}+8\ln 2+\frac {10} {3}+\ln3 
-\ln\left(4\pi^2\beta G^2m^3M^2\right)}
\end{equation}

\section{Specific Heats}

\setcounter{equation}{0}

 We deal  now with a specific heat constructed via 
${\cal C}=\frac {\partial<{\cal U}>} {\partial T}$. Thus 
for $=\frac {3} {2}$ we obtain
\[{\cal C}=\frac {8k[\ln(\pi^2G^2m^3M^2)-4-\ln(kT)-2\boldsymbol{C}]}
{\frac {22} {3}+\ln(kT)-\ln(16\pi^2G^2m^3M^2)}-\]
\begin{equation}
\label{ep7.1}
\frac {8k[3\ln(\pi^2G^2m^3M^2)-3-\ln(kT)-2\boldsymbol{C}]}
{\left[\frac {22} {3}+\ln(kT)-\ln(16\pi^2G^2m^3M^2)\right]^2}
\end{equation}
\noindent 
For $q=\frac {1} {3}$ one has
\[{\cal C}=\frac {9k[11+3\ln 3-\ln 16+3\ln(kT)-3\ln(4\pi^2G^2m^3M^2)-
5\boldsymbol{C}]}
{8\left[\boldsymbol{C}+8\ln 2+3\ln 3+\frac {10} {3}+\ln(kT)-
\ln(\pi^2G^2m^3M^2)\right]}-\]
\begin{equation}
\label{ep7.2}
\frac {9k[8+3\ln 3-\ln 16+3\ln(kT)-3\ln(4\pi^2G^2m^3M^2)-
5\boldsymbol{C}]}
{8\left[\boldsymbol{C}+8\ln 2+3\ln 3+\frac {10} {3}+\ln(kT)-
\ln(\pi^2G^2m^3M^2)\right]^2}
\end{equation}

\nd Figs. 1 - 2 depict specific heats  corresponding
to Eqs.  (\ref{ep7.1}) - (\ref{ep7.2}). We call 
$E=G^2m^3M^2$ with $m<<<M$. We express quantities in $k_BT/E$-units.  Specific heats are negative, as befits gravitation. 
\nd Indeed, such an
occurrence has been associated to self-gravitational systems
\cite{lb}. In turn, Verlinde has associated this type of
systems to an entropic force \cite{verlinde}. It is natural to
conjecture then that such a force may appear at the energy-associated poles. 
 Notice also that temperature ranges are restricted. There is
an $T-$lower bound.

\newpage
\begin{figure}[h]
\begin{center}
\includegraphics[scale=0.6,angle=0]{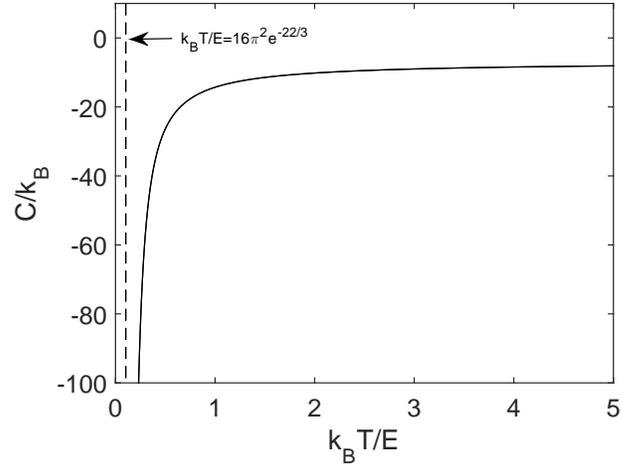}
\vspace{-0.2cm} \caption{Specific heat versus 
$k_BT/E$ for $q=3/2$. It is well known that gravitational
effects make specifics heats to be negative \cite{lb}.
This is clearly appreciated in this graph and in the following one.}\label{fig1}
\end{center}
\end{figure}

\newpage
\begin{figure}[h]
\begin{center}
\includegraphics[scale=0.6,angle=0]{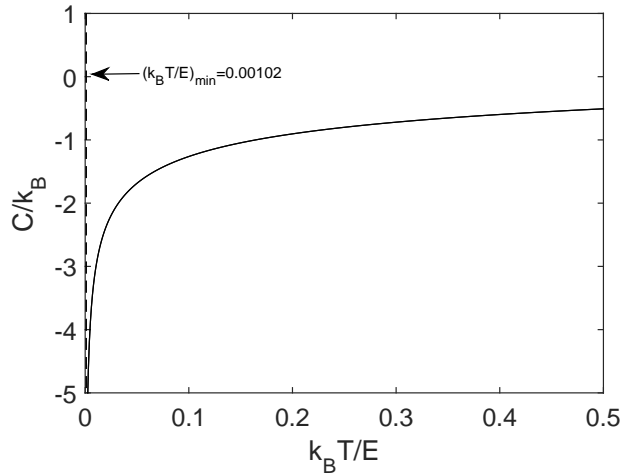}
\vspace{-0.2cm} \caption{Specific heat versus 
$k_BT/E$ for $q=1/3$. It is well known that gravitational
effects make specifics heats to be negative \cite{lb}.
This is clearly appreciated here.}\label{fig2}
\end{center}
\end{figure}

\section{Discussion}

\setcounter{equation}{0}

\nd In this work we have appealed to an elementary regularization
procedure to study the poles in the partition function and the mean
energy that appear, for specific, discrete q-values, in Tsallis'
statistics of Newton's two-body problem. We studied the thermodynamic 
behavior at the poles and
found interesting peculiarities. The analysis was made in one, two,
three, and $3$ dimensions. Amongst the pole-traits we emphasize:

\begin{itemize}

\item The poles appear, both in the partition function and the mean energy, for  
$q\neq 1$

\item These poles ar an artifact of having $q \ne 1$.

\item We have proved that there is a lower bound to the
temperature at the poles.

\item Negative specific heats, characteristic trait of
self-gravitating systems, are encountered.

\end{itemize}

 \vskip 3mm \nd The poles arise only because $q \ne 1$. They are a property of the entropic quantifier, not of the Hamiltonian. Indeed, only for $q \ne 1$ a Gamma function appears in the partition function. It is this Gamma function that displays poles. 

\vskip 3mm \nd Future research should be concerned with cases where it is already known in advance that $q \ne 1$. For these cases, the traits here discovered may acquire some degree of physical     
''reality''.  
\vskip 3mm
\nd In this effort we limit ourselves to the two body problem, as the divergences produced by  gravitation emerge already at the two-body level. The reader is reminded that N-body gravitation is a frontier research topic of Celestial Mechanics.

\nd
The $q=1$ case cannot be analyzed with the present formulation, that is
{\it not valid} for it. The  $q=1$ scenario is discussed in the paper
"Gravitational partition function for the Boltzmann-Gibbs classical distribution"
by D. J. Zamora, M. C. Rocca, A. Plastino and G. L. Ferri.
(see the paper in Researchgate)

\nd The importance of the present communication resides in that fact of having disclosed Tsallis' entropy traits that could not have been suspected before.

\renewcommand{\thesection}{\Alph{section}}

\renewcommand{\theequation}{\Alph{section}.\arabic{equation}}

\setcounter{section}{1}

\section*{Appendix: a simple example of Dimensional Regularization}

\setcounter{equation}{0}

This Appendix illustrates Dimensional Regularization (DR)
with a simple example. The justification of the DR procedure is given in  \cite{tp1}, being a generalization of the treatment advanced in  \cite{tq1}. 

\nd Here we discuss the partition function  ${\cal Z}$ of the 3D Harmonic Oscillator (HO) for
 $q=\frac {2} {3}$. 
We first consider ${\cal Z}$ in en $\nu$ dimensions
\begin{equation}
\label{eqa.1}
{\cal Z}=\int[1+(1-q)\beta(P^2+Q^2)]^{\frac {1} {q-1}}d^\nu pd^\nu q,
\end{equation}
where  $P^2=p_1^2+p_2^2+\cdot\cdot\cdot p_\nu^2$ and
$Q^2=q_1^2+q_2^2+\cdot\cdot\cdot q_\nu^2$.
For the integral we use hyper-spherical coordinates. One finds
\begin{equation}
\label{eqa.2}
{\cal Z}=\frac {2\pi^\nu} {\Gamma(\nu)}
\int\limits_0^{\infty} S^{2\nu-1}  [1+(1-q)\beta S^2]^{\frac {1} {q-1}}dS,
\end{equation}
with $S^2=P^2+Q^2$. Effecting the change $S^2=x$ we obtain
\begin{equation}
\label{eqa.3}
{\cal Z}=\frac {\pi^\nu} {\Gamma(\nu)}
\int\limits_0^{\infty} x^{\nu-1}  [1+(1-q)\beta x]^{\frac {1} {q-1}}dx,
\end{equation}
or
\begin{equation}
\label{eqa.4}
{\cal Z}=\frac {\pi^\nu} {\Gamma(\nu)}
\int\limits_0^{\infty} 
\frac {x^{\nu-1}}  {[1+(1-q)\beta x]^{\frac {1} {1-q}}}dx.
\end{equation}
To evaluate this integral we look it up in  \cite{grad} and find
\begin{equation}
\label{eqa.5}
\frac {\pi^\nu} {\Gamma(\nu)}
\int\limits_0^{\infty} 
\frac {x^{\mu-1}}  {(1+\gamma x)^v}dx=
\gamma^{-\mu}B(\mu,v-\mu),
\end{equation}
where  $B(\mu,v-\mu)$ is Euler's Beta function.
Comparing (\ref{eqa.4}) with  (\ref{eqa.5}) one encounters 
$\mu=\nu$, $v=\frac {1} {1-q}$, $\gamma=(1-q)\beta$, and then

\begin{equation}
\label{eqa.6}
{\cal Z}=\frac {\pi^\nu} {\Gamma(\nu)}[\beta(1-q)]^{-\nu}
B\left(\nu,\frac {1} {1-q}-\nu\right).
\end{equation}
We see that for $q=\frac {2} {3}$ and $\nu=3$ (\ref{eqa.6})
diverges since

\begin{equation}
\label{eqa.7}
B\left(\nu,\frac {1} {1-q}-\nu\right)=
\frac {\Gamma(\nu)\Gamma\left(\frac {1} {1-q}-\nu\right)}
{\Gamma\left(\frac {1} {1-q}\right)},
\end{equation}
with $\Gamma(z)$ being Euler's Gamma function, that exhibits poles   at
 $z=0, -1,-2,-3,....$. From 
(\ref{eqa.7})  it follows that

\begin{equation}
\label{eqa.8}
{\cal Z}=\frac {\pi^\nu} {\Gamma(\nu)}[\beta(1-q)]^{-\nu}
\frac {\Gamma(\nu)\Gamma\left(\frac {1} {1-q}-\nu\right)}
{\Gamma\left(\frac {1} {1-q}\right)},
\end{equation}
or
\begin{equation}
\label{eqa.9}
{\cal Z}=\left[\frac {\pi} {\beta(1-q)}\right]^\nu 
\frac {\Gamma\left(\frac {1} {1-q}-\nu\right)}
{\Gamma\left(\frac {1} {1-q}\right)}.
\end{equation}
Setting  $q=\frac {2} {3}$ in (\ref{eqa.9}) one finds

\begin{equation}
\label{eqa.10}
{\cal Z}=\left(\frac {3\pi} {\beta}\right)^\nu 
\frac {\Gamma\left(3-\nu\right)}
{\Gamma\left(3\right)}.
\end{equation}
Since  $\Gamma(3)=2$, this leads to

\begin{equation}
\label{eqa.11}
{\cal Z}=\frac {1} {2}\left(\frac {3\pi} {\beta}\right)^\nu 
\Gamma\left(3-\nu\right).
\end{equation}
Note that for $\nu=3$, ${\cal Z}$ indeed diverges.  Bollini-Giambiagi's DR approach consists in 
performing the Laurent-expansion of ${\cal Z}$ around  $\nu=3$ and select afterwards, as the physical result for   ${\cal Z}$, the $\nu-3$-independent term in the expansion. The justification for such a procedure is clearly explained in  \cite{tp1}. In order to proceed with the Laurent expansion we first define
\begin{equation}
\label{eqa.12}
f(\nu)=\left(\frac {3\pi} {\beta}\right)^\nu,
\end{equation}
whose Taylor's expansion is
\begin{equation}
\label{eqa.13}
f(\nu)=\left(\frac {3\pi} {\beta}\right)
\sum\limits_{n=0}^{\infty}\ln^n\left(\frac {3\pi} {\beta}\right)
\frac {(\nu-3)^n} {n!}.
\end{equation}
The Gamma function Laurent expansion is

\begin{equation}
\label{eqa.14}
\Gamma(3-\nu)=\frac {1} {3-\nu}+\boldsymbol{C}+
\sum\limits_{m=1}^{\infty}c_m(3-\nu)^m,
\end{equation}
where  $\boldsymbol{C}$ is Euler's constant. 
Multiplying the two series we have
\begin{equation}
\label{eqa.15}
f(\nu)\Gamma(3-\nu)=\left(\frac {3\pi} {\beta}\right)^3\frac {1} {3-\nu}+
\left(\frac {3\pi} {\beta}\right)^3\boldsymbol{C}-
\left(\frac {3\pi} {\beta}\right)^3\ln\left(\frac {3\pi} {\beta}\right)+
\sum\limits_{m=1}^{\infty}a_m(3-\nu)^m.
\end{equation}
Accordingly, ${\cal Z}$    becomes
\begin{equation}
\label{eqa.16}
{\cal Z}=\frac {1} {2}
\left(\frac {3\pi} {\beta}\right)^3\left[\boldsymbol{C}-
\ln\left(\frac {3\pi} {\beta}\right)\right],
\end{equation}
or
\begin{equation}
\label{eqa.17}
{\cal Z}=\frac {1} {2}
\left( 3\pi k_BT\right)^3\left[\boldsymbol{C}-
\ln\left(3\pi k_BT\right)\right].
\end{equation}
Since one demands ${\cal Z}>0$,  $T$ obeys
\begin{equation}
\label{eqa.18}
0<T<\frac {e^{\boldsymbol{C}}} {3\pi k_B},
\end{equation}
entailing an upper bound for $T$, typical of Tsallis' formalism (see \cite{PP94}).

\end{document}